\documentclass[bg]{copernicus}
\usepackage{xcolor}
\usepackage{multicol}               
\usepackage{multirow}
\usepackage[normalem]{ulem}

\usepackage[utf8]{inputenc}
\usepackage[T1]{fontenc}

\usepackage{hyperref}
\definecolor{darkblue}{rgb}{0,0,.6}
\hypersetup{colorlinks=true, breaklinks=true, linkcolor=darkblue, menucolor=darkblue, urlcolor=darkblue, citecolor=darkblue}

\begin{document}

\title{Technical Note: Approximate Bayesian parameterization of a process-based tropical forest model}

\author[1,2]{F. Hartig}
\author[1,3]{C. Dislich}
\author[1]{T. Wiegand}
\author[1]{A. Huth}

\affil[1]{UFZ -- Helmholtz Centre for Environmental Research, Department of Ecological Modelling, Permoserstr. 15,\hack{\newline} 04318 Leipzig, Germany}
\affil[2]{University of Freiburg, Department of Biometry and Environmental System Analysis, Tennenbacher Str. 4,\hack{\newline} 79085 Freiburg, Germany}
\affil[3]{University of G\"{o}ttingen, Department of Ecosystem Modelling, B\"{u}sgenweg 4,\hack{\newline} 37077 G\"{o}ttingen, Germany}

\runningtitle{Approximate Bayesian parameterization of a tropical forest model}

\runningauthor{Hartig et al.}

\correspondence{F.~Hartig (florian.hartig@biom.uni-freiburg.de). Apart from the layout, this preprint corresponds to the final published version, which is available at \href{http://dx.doi.org/10.5194/bg-11-1261-2014}{http://dx.doi.org/10.5194/bg-11-1261-2014}}

\received{28 May 2013}
\pubdiscuss{7 August 2013}
\revised{12 January
2014}
\accepted{17 January 2014}
\published{27 February 2014}

\maketitle

\begin{abstract}
Inverse parameter estimation of process-based models is a long-standing
problem in many scientific disciplines. A key question for inverse parameter
estimation is how to define the metric that quantifies how well model
predictions fit to the data. This metric can be expressed by general cost or
objective functions, but statistical inversion methods require a particular
metric, the probability of observing the data given the model parameters,
known as the likelihood.

For technical and computational reasons, likelihoods for process-based
stochastic models are usually based on general assumptions about variability
in the observed data, and not on the stochasticity generated by the model.
Only in recent years have new methods become available that allow the
generation of likelihoods directly from stochastic simulations. Previous
applications of these approximate Bayesian methods have concentrated on
relatively simple models. Here, we report on the application of a
simulation-based likelihood approximation for FORMIND, a parameter-rich
individual-based model of tropical forest dynamics.

We show that approximate Bayesian inference, based on a parametric likelihood
approximation placed in a conventional Markov chain Monte Carlo (MCMC)
sampler, performs well in retrieving known parameter values from virtual
inventory data generated by the forest model. We analyze the results of the
parameter estimation, examine its sensitivity to the choice and aggregation
of model outputs and observed data (summary statistics), and demonstrate the
application of this method by fitting the FORMIND model to field data from an
Ecuadorian tropical forest. Finally, we discuss how this approach differs
from approximate Bayesian computation (ABC), another method commonly used to
generate simulation-based likelihood approximations.

Our results demonstrate that simulation-based inference, which offers
considerable conceptual advantages over more traditional methods for inverse
parameter estimation, can be successfully applied to process-based models of
high complexity. The methodology is particularly suitable for heterogeneous
and complex data structures and can easily be adjusted to other model types,
including most stochastic population and individual-based models. Our study
therefore provides a blueprint for a fairly general approach to parameter
estimation of stochastic process-based models.
\end{abstract}

\introduction

Parameter estimation of process-based models is a long-standing problem in
many scientific disciplines. Early proponents of process-based modeling in
ecology have stressed the importance of deriving predictions from basic
physical processes, with physical parameters that can be experimentally
determined \citep{Bossel-Real-structureprocessdescription-1992}. In practice,
however, for various reasons ranging from time limitations to fundamental
observability restrictions, most process-based models have parameters for
which direct measurements are not available
\citep{Hartig-Connectingdynamicvegetation-2012}. These parameters need to be
estimated inversely, meaning that they are adjusted by comparing model
outputs to observed data.

To make this comparison, Bayesian methods have become increasingly popular in
ecological research during the last decade
\citep[e.g.,][]{O'Hara-Bayesiananalysisof-2002,
Clark-Whyenvironmentalscientists-2005,
Purves-Environmentalheterogeneitybird-2007,
Higgins-physiologicalanalogyniche-2012}. In addition to their flexibility and
explicit treatment of parameter uncertainty, a particularly appealing
property of Bayesian statistics is that they offer the possibility of
combining existing information on likely parameters values with the
information that is generated inversely
\citep{Hartig-Connectingdynamicvegetation-2012}. As with other inverse
parameterization approaches, Bayesian methods require the definition of a
metric that quantifies how well model predictions fit to the observed data.
In nonstatistical inversion approaches, such metrics are often called goal
functions, objective functions or cost functions
\citep[e.g.,][]{Schroeder-Analysisofpattern-process-2006}. Bayesian approaches
use a particular statistical metric, the probability of obtaining the
observed data given the current model and parameter values, usually referred
to as the likelihood.

Most previous applications of Bayesian statistics to process-based ecological
models derive this probability by making distributional assumptions about how
observations vary around mean model predictions that are independent of the
processes in the model, either ad hoc or based on the observed variance in
the data \citep[e.g.,][]{Martinez-Disentanglingformationof-2011,
Oijen-Bayesiancalibrationcomparison-2013}. This is usually justified with the
idea in mind that there is observation uncertainty or variability in
environmental conditions that is not accounted for in the model. The approach
of constructing likelihoods from such assumptions is the current
state of the art, but it has a major limitation: many process-based
ecological models are already stochastic and predict variability of certain
model outputs. In principle, one would prefer using this variability for
deriving the likelihood, because it is based on our mechanistic understanding
of the ecological system and accounts, for example, for the possibility that
expected variability may change with the parameters of the ecological
processes. Moreover, once we base our likelihood on the outputs of the
stochastic model, additional observer submodels that describe how field data
were collected could easily be added
\citep[e.g.,][]{Zurell-virtualecologistapproach-2009}. However, while
theoretically possible, calculating likelihoods for such complicated
stochastic interactions used to be intractable in practice
\citep{Hartig-Statisticalinferencestochastic-2011}.

This technical limitation has been reduced in recent years by novel
simulation-based approximation techniques that allow practically any
stochastic model to be treated in a formal statistical inference framework.
Of those, approximate Bayesian computation
\citep[ABC;][]{Beaumont-ApproximateBayesianComputation-2010} has arguably
attracted most attention, but there are other approaches as well. Their
common principle is very simple: what is needed for including a stochastic
simulation model in a formal inferential framework is the likelihood
$p(D|M(\phi))$ for an outcome $D$ to occur under a model $M$ with parameters
$\phi$ \citep{Diggle-MonteCarloMethods-1984}. Simulation-based likelihood
approximations estimate this probability by generating draws from the
stochastic model. Subsequently, different methods are used to approximate the
likelihood or posterior \citep{Hartig-Statisticalinferencestochastic-2011}.
Often, this involves comparing the model output and observed data by means of
data aggregations, also called patterns \citep{Wiegand-ExpansionofBrown-2004,
Grimm-Patternorientedmodelling-2012} or summary statistics
\citep{Beaumont-ApproximateBayesianComputation-2010,
Wood-Statisticalinferencenoisy-2010}. For brevity, we will refer to these
methods in general simply as likelihood approximations, or, in the context of
a Bayesian analysis, as approximate Bayesian methods.

The potential of likelihood approximations in ecology has been repeatedly
stressed, but applications to community or population ecology are still rare
\citep[but see][]{Jabot-Inferringparametersof-2009,
Jabot-AnalyzingTropicalForest-2011, May-Metacommunitymainlandisland-2013}. To
our knowledge, there is no previous study that applies likelihood
approximations to a computationally expensive, parameter-rich model
simulating an ecological community.

The aim of our study is to show that simulation-based likelihood
approximations can be successfully applied to complex process-based models.
We use a simulation-based likelihood approximation proposed by
\citet{Wood-Statisticalinferencenoisy-2010} to infer the parameters of
FORMIND, an individual-based model of tropical forest communities. We first
fit to virtual inventory data that were generated from the model with known
parameters. This allows us to test whether the method can correctly identify
all model parameters for different kinds of field data, and to examine how
choice and aggregation (summary statistics) of data affect the results of the
inference. Finally, we apply the method to fit the model to field data from a
tropical montane forest in Ecuador.

\section{Materials and methods}

\begin{table*}[t]
\caption{Important model parameters and their interpretation. The parameter
$dbh_\mathrm{max}$ in the last line refers to the maximum diameter of a tree, which is a
species or PFT-specific parameter of the model
\citep[see][]{Dislich-Simulatingforestdynamics-2009}.}
{\begin{tabular}{lll} \tophline
{Parameter} & {Notes} & Units \\
\middlehline

Light extinction coefficient  & Fraction of light intercepted per unit of LAI       &  [\unit{m^2 m^{-2}}] \\
Recruitment rate                    &  Nnew tree individuals at 1cm dbh        &  [\unit{ha^{-1} yr^{-1}}] \\
Min light for establishment  &      For recruitment, relative to full irradiance    &  [-] \\
Mortality rate    & Probability per year  & [yr$^{-1}$] \\
Falling probability     & Applies to large trees that die & [--] \\
Leaf area index per tree      & Projected leaf area per unit area & [\unit{m^2 m^{-2}}] \\
Crown diameter    & Relative to diameter  & [--] \\
Crown length      & Relative to tree height     & [--] \\
Max dbh growth (gro$_{\max}$)&  Maximum stem diameter (sd) growth rate & [mm\,yr$^{-1}$] \\
Start growth      &  sd growth rate for minimum stem diameter (relative to gro$_{\min}$)& [--]   \\
End growth  &     sd growth rate at maximum stem diameter (relative to gro$_{\max}$)     & [--]  \\
Max growth diameter & Stem diameter of max growth (relative to $dbh_\mathrm{max}$) & [--]  \\ \bottomhline \\
\end{tabular}}
\label{table: model parameters}
\end{table*}

\subsection{Forest gap dynamics and the FORMIND model}

Forest ecosystems are locally highly dynamic. One of the most prominent
drivers of these dynamics, particularly in the tropics, are natural
disturbances, where large trees that have lost stability due to mortality or
other factors fall and damage or kill other trees. Gap formation creates a
dynamic mosaic of light-filled gaps in natural forests
\citep[e.g.,][]{Shugart-TheoryForestDynamics-1984,
McCarthy-Gapdynamicsforest-2001}. Within these gaps, pioneer species colonize
first, until other species take over and continue the successional dynamics
that are thought to be one part of the explanation for forest diversity
\citep[e.g.,][]{Kohyama-Size-StructuredTreePopulations-1993}.

Mechanistic forest models that describe the processes of gap formation and
recovery have a long history in ecology
\citep{Pacala-ForestModelsDefined-1996, Bugmann-SimplifiedForestModel-1996,
Shugart-Terrestrialecosystemsin-1998, Huth-Simulationofgrowth-2000}. These
models typically include several tree species with different growth
properties and light demands. For highly diverse systems such as tropical
rainforests, species are usually grouped into plant functional types (PFTs)
that represent a group of species with similar functional properties.
Parameters and model predictions per plant functional types then represent a
mean over the species that are represented by this type. Gap formation by
falling dead trees maintains the modeled forest in a dynamic equilibrium. As
a result, forest gap models do not merely predict a mean value for outputs
such as biomass, species composition, or tree size distributions. Rather,
they deliver samples of different possible values for these outputs and
therefore allow probabilities to be assigned to different community or
biomass states. These predictions of spatiotemporal variation in community
composition is what we will use later to derive a probabilistic measure of
distance between the model output and observed data.

FORMIND, the forest model used for this study, is a stochastic,
individual-based forest model designed in the tradition of classical forest
gap models \citep{Koehler-Modellinganthropogenicimpacts-2000}. It has been
applied for estimating forest succession, variability and disturbances
impacts in various tropical locations around the world
\citep[e.g.,][]{Rueger-Ecologicalimpactsdifferent-2007,
Kohler-Towardsground-truthingof-2010, Dislich-Modellingimpactshallow-2012,
Gutierrez-Successionalstagesprimary-2012}. The simulation area (plot) in
FORMIND, which can be of variable size (we use 1~ha throughout the paper) is
subdivided into 20\,m$\times20$\,m grid cells. Tree individuals are assigned to
one of these cells and interact with each other on the cell, but do not have
an explicit spatial position within the cells. The model state is entirely
described by species or functional type, size (measured in diameter
at breast height $dbh$), and location (cell) of all trees. Other variables,
such as tree height and crown dimensions, are derived through fixed
allometric relationships.

At each time step (we use 5\,yr time steps), the light climate in each cell
is calculated from the trees on that cell and their respective crowns.
Subsequently, establishment (light-dependent, stochastic), mortality
(stochastic) and tree growth (light-dependent) act on all tree individuals.
Important parameters in the model (Table~\ref{table: model parameters}) are
recruitment and mortality rates, parameters that describe the size-specific
maximum growth rates, and the allometric relationships that determine height
and crown dimensions. Details of these processes, together with a more
detailed description of the model scheduling, are provided in the Supplement
\citep[see also][]{Koehler-Modellinganthropogenicimpacts-2000,
Dislich-Simulatingforestdynamics-2009}.

\begin{figure*}[t]
\includegraphics[width=15cm]{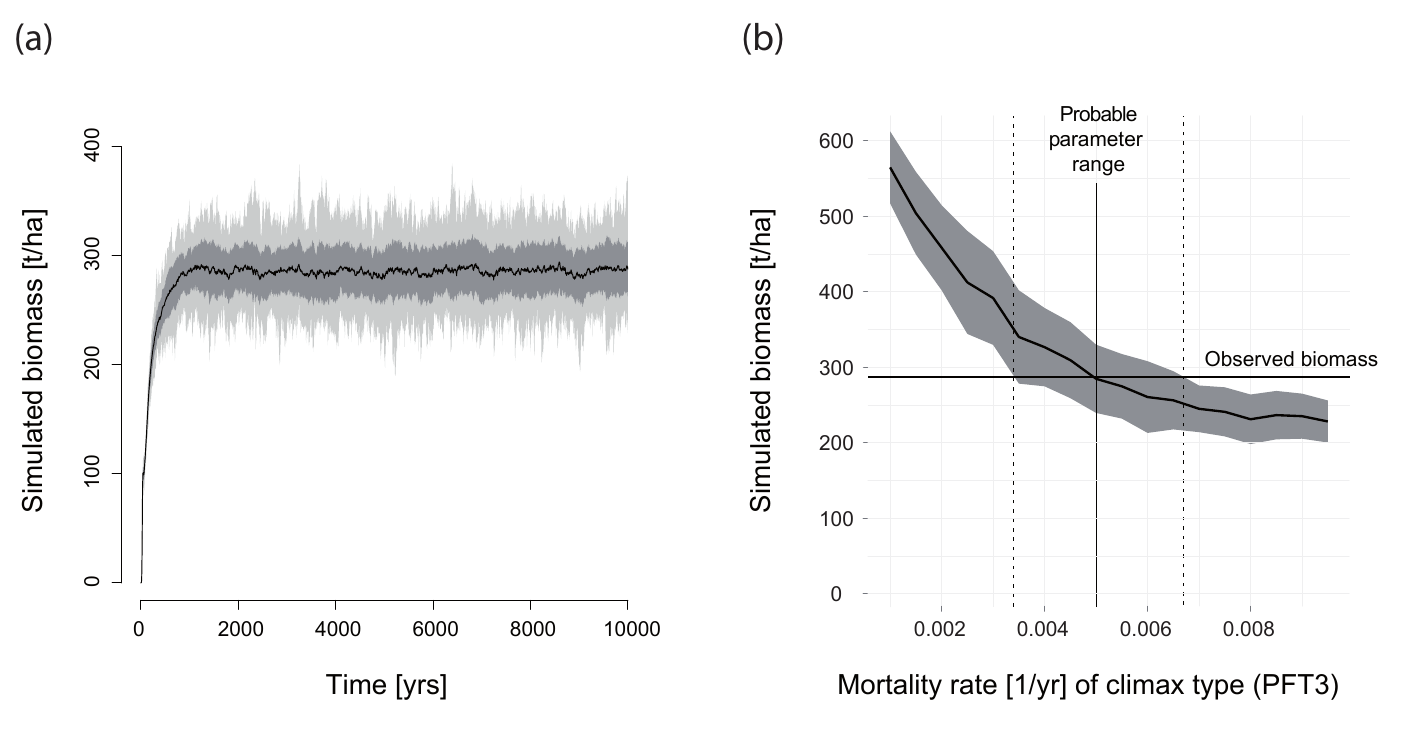}
\caption{Principle of statistical inference through stochastic simulation.
 \textbf{(a)} shows mean model predictions (black), standard deviation
(gray) and min/max values (light gray) for the biomass of a 1\,ha plot over
10\,000\,yr, starting from an empty plot. \textbf{(b)} shows the same
mean equilibrium biomass (black) and two standard deviations (gray), but as a
function of the mortality of the late-successional type PFT 3; all other
parameters constant. Comparing the observed biomass from  \textbf{(a)},
which was created with a mortality rate of 0.005, with the predicted biomass
for different mortality rates, we can infer the original value as well as a
statistical uncertainty, without having to define a statistical model.
\label{figure: Model stochasticity}}
\end{figure*}

\subsection{Bayesian parameter estimation with\hack{\newline} simulation-based likelihood approximations}

We use a Bayesian approach for parameter estimation. One of the advantages of
using Bayesian methods with Markov chain Monte Carlo (MCMC) sampling for
simulation-based likelihood approximations is that MCMCs, unlike optimization
approaches, are more robust towards variance in likelihood estimates
generated by the approximation
\citep{Hartig-Statisticalinferencestochastic-2011}. Bayesian methods are also
somewhat better suited to dealing with interactions between parameters, which
is a phenomenon to be expected in process-based models. In principle,
however, one could use the likelihood approximation used in this study with
an optimization algorithm in a maximum-likelihood framework as well.

There are a number of introductions to Bayesian statistics. A detailed
reference is \citet{Gelman-BayesianDataAnalysis-2003}; for a shorter
introduction see \citet{Ellison-Bayesianinferencein-2004}. We give only a
brief summary here. The outcome of a Bayesian inference is a probability
distribution $P(\phi|D_\mathrm{obs})$ for the parameters $\phi$ given the
observed data $D_\mathrm{obs}$. This distribution, called the posterior, is
calculated as
\begin{equation}\label{eq: bayes formula}
      p(\phi|D_\mathrm{obs}) = c \cdot p(D_\mathrm{obs}|M(\phi)) \cdot p(\phi) \;,
\end{equation}
where $c$ is a normalization constant, the prior probability density
$p(\phi)$ quantifies parameter uncertainties before comparing the model to
the observed data, and the likelihood function $p(D_\mathrm{obs}|M(\phi))$
describes the probability of obtaining the observed data conditional on the
model M with parameters $\phi$. Broadly speaking, we may say the likelihood
quantifies the quality of the fit, while the prior quantifies our prior
expectation for each possible parameter value.

Because our main concern in this paper is the approximation of the
likelihood, we chose wide uniform (flat) priors for all parameters and data
types, which means that the posterior and likelihood are strictly
proportional to each other across the possible prior range. Tables with the
widths of these uniform priors are provided in the Supplement. Given that we
knew that the model reacts nonlinearly to many parameters, other
uninformative prior choices would have been possible
\citep[e.g.,][]{Kass-selectionofprior-1996}, but we felt for the purpose of
our study it is more useful to ensure proportionality of likelihood and
posterior to facilitate the interpretation of the results.

\subsubsection{Generating approximate likelihoods}

The technical key novelty in this study is the definition of the likelihood
$p(D_\mathrm{obs}|M(\phi))$. In ``conventional'' Bayesian or maximum
likelihood studies, this conditional probability is obtained by formulating
an error model that quantifies probabilities of deviations between model
predictions and observations occurring
\citep[e.g.,][]{VanOijen-Bayesiancalibrationof-2005}. This model may be
mechanistically motivated, for example by knowledge about measurement
uncertainties. In practical situations, however, there are usually a number
of error sources that interact, and error models are therefore typically
either fixed ad hoc \citep{Oijen-Bayesiancalibrationcomparison-2013} or
derived from the observed variability in the data
\citep{Martinez-Disentanglingformationof-2011}. Hence, conventional
likelihoods are usually independent of the mechanisms in the process model
that is fit.

\begin{table*}[t]
\caption{Overview of parameter estimations with different models, parameters
and summary statistics. Abbreviations for the data: SSD = stem size
distribution (16 10\,cm classes); GRO = mean stem diameter growth for each of
the 16 10\,cm stem diameter classes; BM = biomass. If not stated otherwise,
the data type was available for each PFT separately. If we use the mean over
all PFTs, we label this ``total''. ``Full parameters'' means that all
parameters listed in Tables~1 and 2 of the Supplement are estimated inversely.
``Reduced parameters'' means that only recruitment, mortality, maximum growth and
maximum growth diameter are estimated. ``Number of parameters'' and ``data
dimension'' give the number of parameters and data points, respectively.
``Posterior width'' measures the posterior width of the marginal
distributions by the ratio between marginal posterior standard deviation and
uniform prior width averaged over all parameters. ``Convergence ranking''
provides a ranking of the speed of convergence of the MCMCs based on the
convergence diagnostics discussed in the Supplement. Lower numbers indicate
fastest convergence. As E1 uses different data and a different number of
PFTs, the convergence ranking is not fully comparable and was set in
parentheses.}
 \begin{tabular}{llcccc} \tophline
  \multirow{2}{*}{{Label}} & \multirow{2}{*}{{Description}} & {Number of} & {Data} & {Posterior} & {Convergence}\\
 &  & {parameters} & {dimension} &{width} & {ranking} \\ \middlehline
  \multicolumn{3}{l}{{Parameterization to virtual tropical forest (3 PFTs):}} & & &\\
  $V1$ & Field data: SSD, GRO, reduced parameters &  $12$ & $96$ & 0.019 & 1\\
  $V2$ & Field data: SSD, GRO, full parameters &  $26$ & $96$ & 0.096 & 5 \\
  $V3$ & Field data: SSD, reduced parameters &  $12$ & $48$ & 0.073 & 2 \\
  $V4$ & Field data: total SSD, reduced parameters &  $12$ & $16$ & 0.190 & 3 \\
  $V5$ & Field data: BM, reduced parameters &  $12$ & $3$ & 0.192 & 4 \\ [0.2cm]
  \multicolumn{3}{l}{{Parameterization to Ecuadorian montane rain forest (7 PFTs):}} & & & \\
  $E1$ & Field data: SSD  &  $18$ & $112$ & 0.036 & (5) \\ \bottomhline \\
\end{tabular}
\label{Table: Fit types}
\end{table*}

Our approach goes beyond such an independent error model towards an approach
where both the mean model prediction and the probability of observing
deviations from the mean are derived from the same stochastic ecological
processes. This is particularly promising in systems where
process stochasticity dominates observation errors. For inventory data from
tropical forests, this is generally the case. Given typical observation
errors \citep[see][]{Chave-Errorpropagationand-2004}, we can assume that, for
small plots, observation uncertainty is small compared to local biomass
variation due to successional dynamics
\citep[e.g.,][]{Chave-Spatialandtemporal-2003}. FORMIND simulations of the
aforementioned successional dynamics triggered by gap formation explain the
extent of this variability well
\citep[e.g.,][]{Kohler-Towardsground-truthingof-2010} and can therefore be
used to generate statistical expectations for model outputs such as biomass
conditional on the model parameters (Fig.~\ref{figure: Model stochasticity}).

Several techniques have been suggested for achieving likelihood
approximations from stochastic simulations. Most prominent is arguably the
method of ABC \citep{Csillery-ApproximateBayesianComputation-2010,
Beaumont-ApproximateBayesianComputation-2010}, which has attracted much
attention in recent years. However, as discussed in
\citet{Hartig-Statisticalinferencestochastic-2011}, there are a number of
closely related methods that are currently not counted as examples of ABC,
but that apply similar principles. In this study, we use the method of
``synthetic likelihoods'' suggested by
\citet{Wood-Statisticalinferencenoisy-2010}, classified as a ``parametric
likelihood approximation'' in
\citet{Hartig-Statisticalinferencestochastic-2011}.

\begin{figure*}[t]
\centering
\includegraphics[width=13cm]{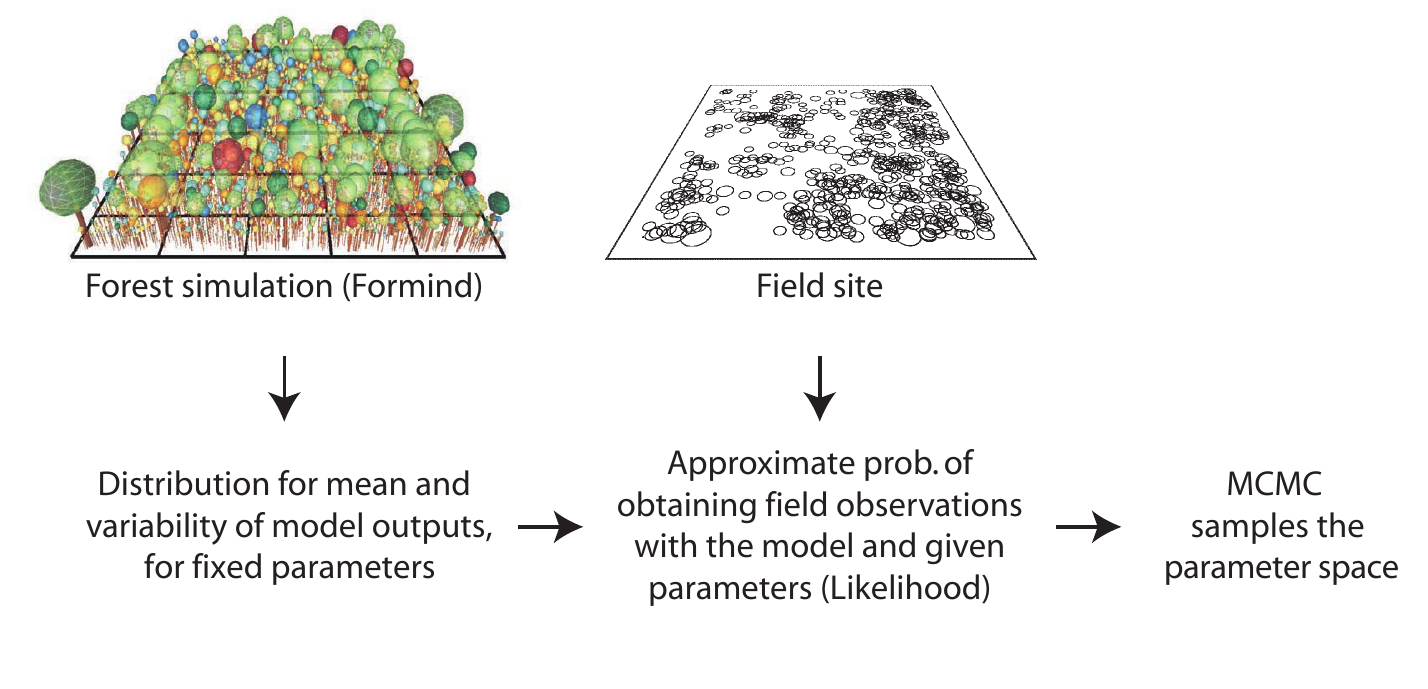}
\caption{Illustration of the estimation process: at the top left, a
visualization representing the FORMIND model. Different colors represent
different PFTs. The model is compared to the field data (middle) by fitting a
distribution to the stochastic model output, and calculating the approximate
probability of observing the field data from this distribution. This
approximate likelihood value feeds into the conventional Bayesian analysis.
\label{figure: analysis chain}}
\end{figure*}

The principle of this method is to estimate $p(D_\mathrm{obs}|M(\phi))$, for
any $\phi$ desired, by fitting a parametric distribution to the output of the
stochastic simulation, and estimating the probability of obtaining
$D_\mathrm{obs}$ from this distribution (Fig.~\ref{figure: analysis chain}).
We used a multivariate normal distribution because it fitted well to the
simulation outputs, and allows a convenient estimation of the covariance
structure, but normality is by no means a fundamental requirement of the
approach. For the multivariate normal approximation, the likelihood of
obtaining the observed data $D_\mathrm{obs}$ with model $M$ and parameters
$\phi$ is
\begin{equation}\label{eq: auxiliary function}
\begin{split}
      p(D_\mathrm{obs}|M(\phi)) & \approx c \cdot |\Sigma_\mathrm{sim}(\phi)|^{- 1/2} \exp[- 1/2  \\
      \;\; (D_\mathrm{obs} & - \bar{d}_\mathrm{sim}(\phi))^T \Sigma_\mathrm{sim}^{-1}(\phi) (D_\mathrm{obs} -
      \bar{d}_\mathrm{sim}(\phi))].
\end {split}
\end{equation}
Here, $c = (2 \pi) ^{- k/2}$, with $k$ being the dimension of
$D_\mathrm{obs}$; $\bar{d}_\mathrm{sim}(\phi)$ is the corresponding vector of mean simulation
outputs; $\Sigma_\mathrm{sim}(\phi)$ is the covariance matrix of the simulation
outputs that summarizes variability of and correlations between simulation
outputs; and $|\Sigma_\mathrm{sim}(\phi)|$ is the determinant of the covariance
matrix. Pseudocode for the entire parameter estimation algorithm is provided
in the Supplement.

\subsubsection{Representation of the data}

As in ABC, it is desirable to represent the data used in
Eq.~(2) in a low-dimensional form so that the
estimation particularly of $\Sigma_\mathrm{sim}^{-1}(\phi)$ can be achieved
in a computationally efficient way. The challenge here is to find
lower-dimensional aggregations (summary statistics) of the data that still
contain the same amount of information for the purpose of the inference as
the raw data (sufficiency). Unfortunately, there is still no generally
accepted rule on how to find good summary statistics \citep[but
see][]{Fearnhead-ConstructingSummaryStatistics-2012,
Blum-comparativereviewdimension-2013}. We therefore decided to use mainly two
aggregations that have been frequently used for summarizing inventory data in
forest modeling, and tested their information content by fitting the model to
simulated data. The first aggregation is using stem size distributions, which
count the number of tree individuals per (in our study $10$\,cm) size class
per PFT or for all trees. The second is the size-specific mean growth, which
quantifies the mean stem diameter growth for different size classes. We also
experimented with other forest attributes or aggregations of the data (see
Table~\ref{Table: Fit types}).

\subsubsection{Posterior estimation}

Subsequent posterior estimation based on the approximate likelihood was done
with an adaptive Metropolis--Hastings MCMC
\citep{Haario-adaptiveMetropolisalgorithm-2001}. We always ran several chains
and checked convergence visually and with Gelman--Rubin diagnostics
(\citealp{Gelman-Inferencefromiterative-1992}; see Supplement for further details). As several seconds were
typically required to evaluate a single parameter combination with FORMIND,
posterior estimations cost substantial computing time. The exact number,
length and burn-in of chains are provided in the figure captions of the
Supplement. Figure~\ref{figure: analysis chain} provides a visual summary of
the analysis method.

\subsection{Field data, model setup and analysis}

We used two data sets to fit the parameters of the model, a ``virtual'' 1\,ha
inventory with three plant functional types (PFT1: pioneer; PFT2:
midsuccessional; PFT3: late successional) that was created from the FORMIND
model itself (which has the advantage that the ``true'' parameter values are
known), and a 5\,ha forest inventory from a montane tropical rainforest in
Ecuador that is described in \citet{Dislich-Simulatingforestdynamics-2009}.
The purpose of the virtual data set is to test the parameter estimation method
for different data types in a situation where true parameters are known,
while the data from Ecuador provide a realistic case study that allows us to
test the method in a situation that had previously been dealt with by manual
calibration based on visual assessment of model fit.

To create the virtual inventory, we used a base parameterization that was
adjusted for exhibiting biomass values and successional patterns typical to a
wet tropical lowland rainforest. With this setting, we simulated 1000 model
runs, and created virtual data sets from the mean equilibrium values of these
replicates for different types of output variables (summary statistics) such
as biomass, stem diameter growth rates and stem size distributions. We also
experimented with a different number of parameters to be estimated. A summary
of these options, labeled V1--V5, is provided in Table~\ref{Table: Fit types}.
For complex models, it can usually not be known a priori which data types are
sufficient for a particular inferential question, and we therefore have to
test this with virtual data \citep[see
also][]{Jabot-Inferringparametersof-2009}. The number of estimated
parameters, on the other hand, is more a practical issue: from our
understanding of the processes, it was foreseeable that FORMIND would exhibit
interactions between parameters with respect to these outputs, but it is of
practical interest to determine to which extent posterior estimation is
slowed down by these interactions, and how those interactions look exactly.

\begin{figure*}[t]
\includegraphics [width=17cm]{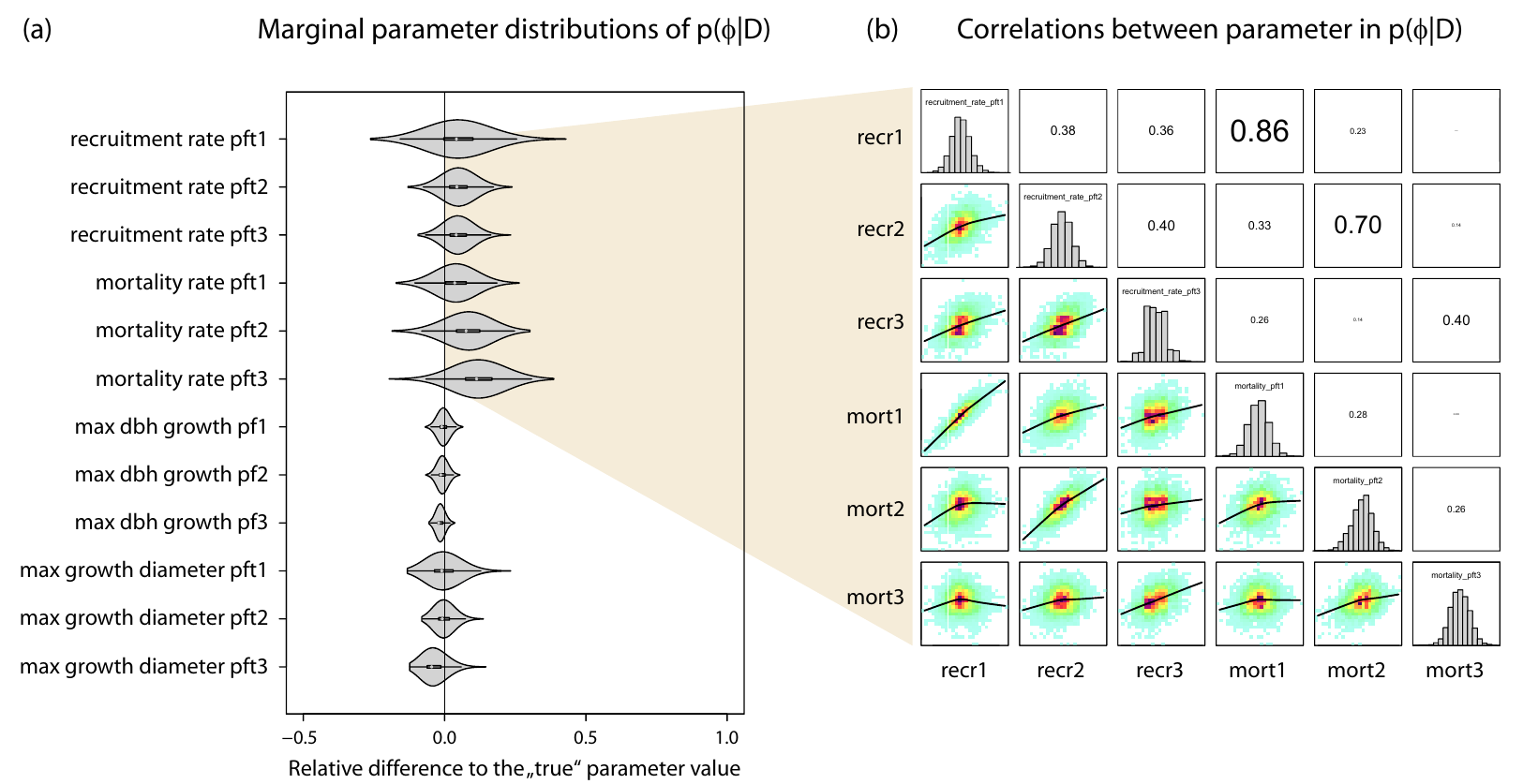}
\caption{Summaries of the estimated parameter values (shown as probability
distributions) after fitting the model to the virtual inventory data (case V1
in Table~\ref{Table: Fit types}). The distributions in \textbf{(a)}
correspond to the marginal posterior density $p(\phi|D)$ for each parameter,
scaled relative to the ``true'' values that were used to create the synthetic
data (see Table~2 in the Supplement for true values and units). The dot
within each distribution denotes the median value. Panels in \textbf{(b)}
visualize correlations between recruitment and mortality parameters in the
posterior sample (recr1 refers to the recruitment rate of PFT1, mort2 refers
to mortality of PFT2 and so on). The diagonal shows the marginal
distributions displayed in panel \textbf{(a)}. The lower triangle shows the
correlation density between the parameters on the diagonal (red values
denoting higher density) and a nonlinear fit of the correlation (black line).
The upper triangle shows Spearman's rank correlation coefficients for the
correlations in the lower triangle. \label{fig: relative bias}}
\end{figure*}

For fitting the model to field data in Ecuador, a tree-species grouping into
seven PFTs was used that is described in detail in
\citet{Dislich-Simulatingforestdynamics-2009}. Due to data availability, we
used only the stem size distributions for the parameter estimation, which we
label E1.

\section{Results}

\subsection{Fit to virtual inventory data (tropical lowland\hack{\newline} rainforest, V1--V5)}

As explained above, we considered a number of options to fit the model to the
virtual inventory data. Those options differed in the aggregation of model
outputs, and in the number of estimated parameters. We concentrate here on
the case V1 in Table~\ref{Table: Fit types} (detailed data, not all
parameters under calibration). Results for the other cases are discussed in
brief below. Detailed results are provided in the Supplement.

\subsubsection{Marginal distributions}

Figure~\ref{fig: relative bias}a shows the estimated marginal posterior
densities (Eq.~\ref{eq: bayes formula}) for the parameterization V1 in
Table~\ref{Table: Fit types}. Those marginal posterior densities represent
the probability assigned for the values of each parameter. We find that most
parameter values are retrieved correctly and show moderate uncertainties on
the order of $20 - 50$\,\% of the mean. When interpreting these plots, note
that ``marginal'' means that we display the values of one particular
parameter in the posterior sample without taking the corresponding values of
the other parameters into consideration. If there are correlations between
parameters densities in the posterior, marginal uncertainties often appear
substantially larger than they effectively are when viewed multivariately. We
examine this in the next subsection.

\subsubsection{Correlations}

Marginal distributions represent a cross-section of the posterior sample
along one parameter, which neglects potential trade-offs between parameters
with respect to the data to which the model is fit. Statistical models are
usually designed to avoid such correlations wherever possible. For
process-based models, on the other hand, the correspondence to specific
biological mechanisms is usually the main design criterion. It is therefore
likely that such correlations will appear when estimating their parameters,
as evidenced by Fig.~\ref{fig: relative bias}b. Moreover, it is to be
expected that the correlation structure depends on the data used to fit the
model. Less informative data will typically lead to more parameter
combinations that can reproduce this data, affecting the correlation
structure in the posterior sample.

These expectations are largely confirmed by our results. We find strong
positive correlations particularly for recruitment and mortality of early
successional types, as one would expect, because, for those PFTs, increased
mortality can be compensated for to some extent by increased recruitment.
Also, we find that the correlation structure changes with the data types
used. A detailed analysis of the correlation structure for the different data
types (summary statistics) tested by us is provided in the Supplement.

\subsubsection{Choice of data type and number of fitted\hack{\newline} parameters}

For the parameter estimation scenarios V3--V5 (Table~\ref{Table: Fit types})
that used less information (more aggregated model outputs or summary
statistics), posterior parameter estimates were wider than for our baseline
scenario V1 (Table~\ref{Table: Fit types}; for details see Supplement). It
can therefore be concluded that all further aggregations of the data used in
V1 lose information for the purpose of estimating the considered parameters
\citep[see also][]{Wiegand-ExpansionofBrown-2004}. Similarly, increasing the
number of fitted parameters (scenario V2) increased the width of the
posterior distribution. For all cases, the results indicate that more coarse
aggregations or more parameters under calibration lead to additional
correlations between parameters with respect to the objective of reproducing
the respective data type, leading to wider marginal distributions and slower
convergence of the MCMC algorithms (overview in Table~\ref{Table: Fit types};
see Supplement for details).

\hack{\newpage}

\begin{figure}[t]
\includegraphics [width=8.5cm]{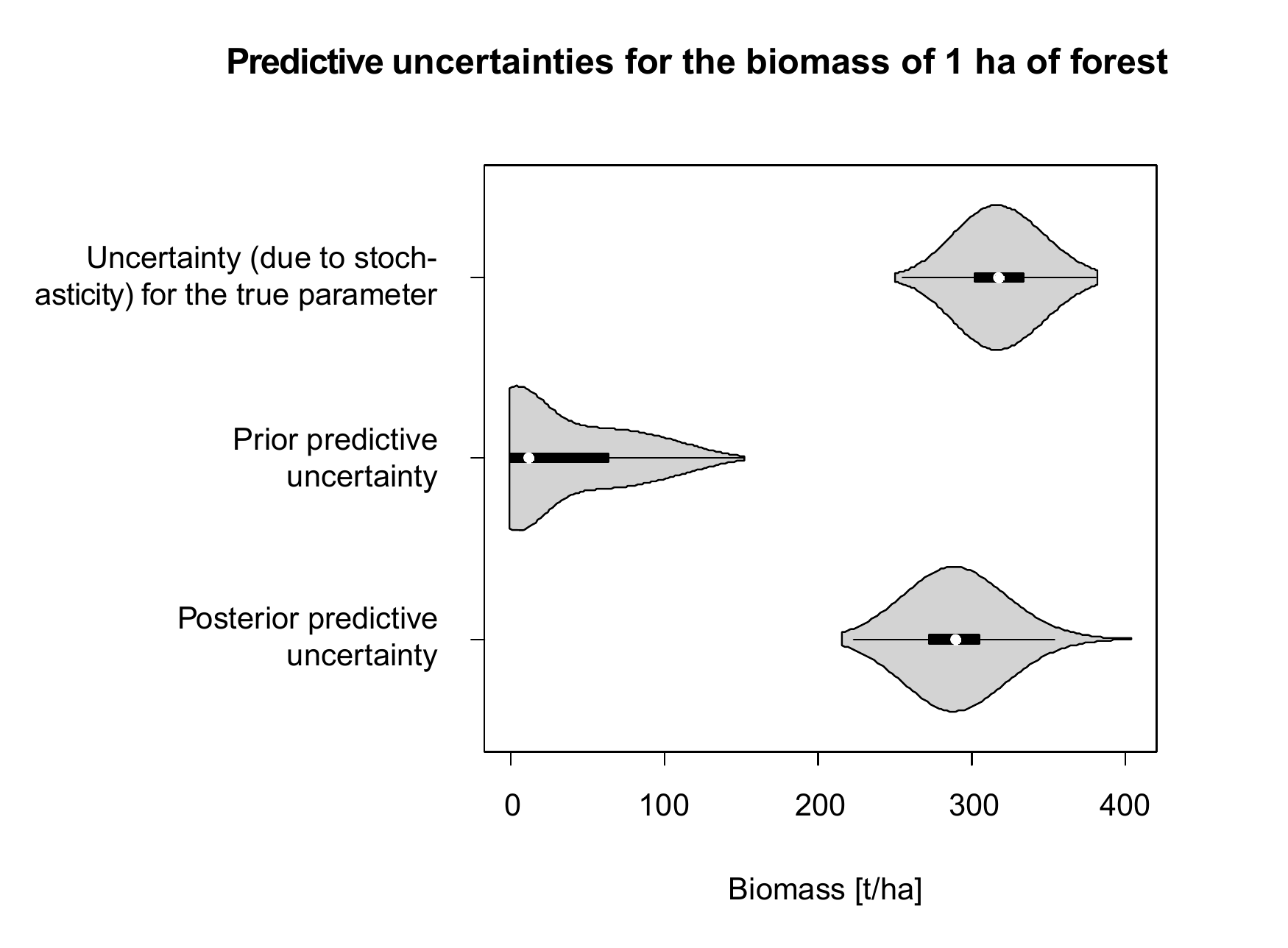}
\caption{True, prior and posterior predictive uncertainty. Each distribution
is created from 1000 model runs, observing the biomass on a 1\,ha forest plot
after 2000 years. The upper distribution shows biomass values from model runs
with the same, ``true'' parameters (Table~2, Supplement), and thereby gives
an estimate of the stochastic uncertainty of the model. For the middle
distribution, model parameters were drawn from the prior distribution
(resulting in what is called the prior predictive distribution). For the
lower distribution, model parameters were drawn from the posterior (posterior
predictive uncertainty). \label{fig: posterior uncertainty}}
\end{figure}

\begin{figure*}[t]
\includegraphics [width=17cm]{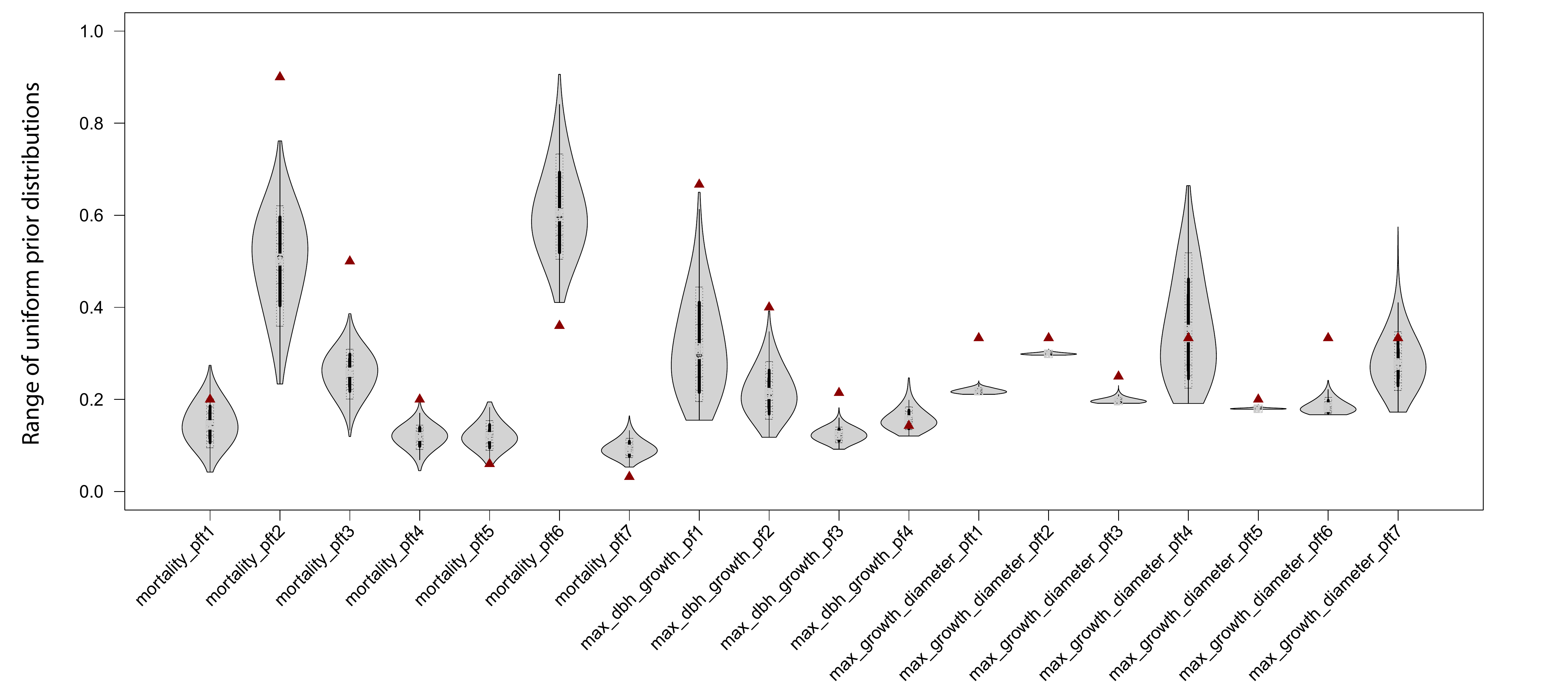}
\caption{Marginal posterior probabilities for the model parameters after
fitting the model to field data from Ecuador, scaled relative to the uniform
prior distributions (see Table~3 in the Supplement for prior values and
units). Values used by \citet{Dislich-Simulatingforestdynamics-2009} are
marked as dark-red triangles. An unscaled version of these distributions and
correlations are provided in Figs.~12 and 13 of the Supplement. \label{fig:
fit ecuador}}
\end{figure*}

\subsubsection{Reduction of predictive uncertainty}

The Bayesian framework also allows convenient estimation of the predictive
uncertainty before and after fitting the model to the data. We compare three
cases, the inherent stochastic uncertainty of the model with the true
parameters, the uncertainty resulting from parameters drawn from the prior
distribution (i.e., before parameter estimation), and the uncertainty for
parameters drawn from the posterior distribution (i.e., after parameter
estimation). The results displayed in Fig.~\ref{fig: posterior uncertainty}
show that the posterior predictive mean is similar to that of the true
parameters, with predictive uncertainty only slightly larger than for the
true, fixed parameter value, which indicates that, for a single 1~ha
plot, the output uncertainty generated from process stochasticity is on the
same order of magnitude as the uncertainty originating from the parameters.
The prior predictive distribution, showing the predictions before
calibration, is biased towards smaller values. This is likely due to the fact
that many parameters in the prior distribution, in particular those with high
mortality, result in very low biomass values.

\subsection{Fit to Ecuadorian montane rain forest, E1}

The results of the fit to field data from Ecuador (case E1 in
Table~\ref{Table: Fit types}) are displayed in Fig.~\ref{fig: fit ecuador}.
We show the marginal distributions for each parameter scaled to the prior
range. Priors were uniform distributions within plausible ranges for a forest
of that type. Hence, the figure provides a visual estimate of the reduction
of parameter uncertainty that would be reached starting from a state at which
no specific information about the plot is available.  A distribution of a
width 0.2, for example, would indicate that the prior uncertainty is reduced by
$80$\,\% with the chosen data type. Parameter correlations and unscaled
marginal parameter estimates are provided in the Supplement, Figs.~13 and
~12, respectively.

\section{Discussion}

Inverse parameter estimation of ecological models requires a metric that
quantifies how well model predictions fit to observed data. Because of
technical limitations, the current state of the art is choosing these metrics
from expert knowledge or deriving them from field data. However, new
statistical methods make it possible to generate goodness-of-fit metrics
directly from any stochastic simulation model. More specifically,
simulation-based likelihood approximations allow the generation of
approximate likelihood functions that return the probability of obtaining a
certain field observation given the model parameters directly from the
stochastic model outputs. This technique provides a universal and unambiguous
way to connect stochastic ecological models to field data.

The present study is one of the first to apply this method to a
parameter-rich ecological model. We use a parametric likelihood
approximation, proposed by \citet{Wood-Statisticalinferencenoisy-2010}, for
fitting FORMIND, a relatively complex individual-based forest gap model, to a
range of different virtual inventory data created from the model as well as
to real field data from an Ecuadorian tropical forest.

\subsection{Validation of the method with virtual inventory data}

Fitting the model to different virtual inventory data sets allowed us to
assess uncertainty and bias of the fit for situations where true parameters
were known. For the most detailed data (abundance and growth distributions,
scenario~V1 in Table~\ref{Table: Fit types}), estimated parameter values were
largely unbiased, with correlations between a few of the parameters
(Fig.~\ref{fig: posterior uncertainty}, as well as Figs.~2 and 3 in the
Supplement). With increasing level of aggregation (scenarios V3--V5) parameter
values showed increasing correlations, bias and uncertainty. Correlations in
the posterior indicate a trade-off between parameters for the purpose of
reproducing the data used for the fit. For scenario V1 (full data), for
example, correlations occurred mostly
between mortality and recruitment of the same PFT, which indicates that higher mortality can
to some extent be compensated by higher recruitment to produce similar
population sizes (as V1 contained growth data, growth rates are tightly
constrained, so the only option to maintain similar population sizes is to
increase recruitment). In scenario V3, which did not include growth data,
posterior parameter estimates also show correlation between mortality and
growth, evidently because growth is unconstrained for this data type.

It is important to take correlations into account when interpreting marginal
parameter uncertainties such as Fig.~\ref{fig: relative bias}a: if there are
correlations between parameters, marginal uncertainties appear wider than in
the multivariate correlation plots. This remains true for higher-order
correlations, which are likely present for more aggregated data types used in
scenarios V4 and V5, but which are difficult to visualize. Comparing the
extent to which model parameters are constrained by the data based only on the
width of their marginal posterior distribution can therefore be
misleading in the presence of strong correlations. It is an advantage of the
Bayesian analysis (or rather the use of an MCMC) that these interactions can
be made explicit and interpreted. Thinking about the reasons for correlations
may also be helpful for understanding and improving the model structure,
although we stress that a correlation in the posterior does not necessarily
mean that a parameter is redundant. It merely means that changes in one of
the parameters may be counterbalanced by the other to maintain the same value
of the model output under consideration. For example, correlations and bias
increase from V1 to V3, indicating that even for fitting recruitment,
mortality and growth parameters only, static data such as stem size
distributions do not provide sufficient information to constrain all
parameters at once. Thus, correlations are connected to a particular data
type, and they inform us as to which parameters cannot be fully constraint by this
data type.

Bias and correlations observed in the scenarios V1--V5 using the virtual
inventory data seemed to originate predominantly from data limitations and
not from problems with the simulation-based likelihood approximation. We saw
no indications that would suggest that the parametric model (multivariate
normal) used in the likelihood approximation created any problems or bias by
not adequately summarizing model outputs, which would be theoretically
possible. However, due to the computational complexity of our study, it was
not possible to make a more systematic analysis of this question, for example
by using virtual replicates of the field data sets or less aggregated data
types.

\subsection{Fit to Ecuadorian field data}

Only static data were available to us for fitting the FORMIND model to field
data from a montane forest in Ecuador. Our previous analysis suggested that
these data would not be sufficient to sensibly constrain all demographic
parameters at once. To get ecologically interpretable results, we therefore
fixed the recruitment parameters to the values used in
\citet{Dislich-Simulatingforestdynamics-2009}, and calibrated mortality and
growth parameters only. Prior uncertainty was considerably reduced by these
data (Fig.~\ref{fig: fit ecuador}), suggesting that our approach together
with the Ecuadorian data is able to substantially constrain the parameters
under calibration. Marginal posterior parameter estimates are similar to
those derived by \citet{Dislich-Simulatingforestdynamics-2009} with a
combination of literature data, expert knowledge and calibration (see
Supplement, Table~3 for exact values).

From the fits to the virtual inventory data V3 (Fig.~7, in Supplement), we
expected correlations in the posterior mostly to occur between parameters of
the same PFT. We find those correlations, but we also find additional
correlations, particularly between the mortality parameters of some PFTs
(Fig.~13, Supplement). To understand this, one has to know that species
grouping designed by \citet{Dislich-Simulatingforestdynamics-2009} is
hierarchical, consisting of 7 PFTs that were further divided into 4 growth
groups with equal maximum diameter growth for the PFTs in each group, with
the following relation between (PFT) and growth group:
(1)--2, (2)--1, (3,4)--3, and (5,6,7)--4. Diameter growth parameter 3, which is
estimated lower, thus applies to the midsuccessional PFTs 3 and 4, and
diameter growth parameter 4, estimated higher, applies to the late
successional PFTs 5, 6 and 7. This hierarchical species grouping is mirrored
in the correlation structure, with particularly strong correlations in the
mortality parameters of PFTs that belong to the same growth group. Our
interpretation of this pattern is that PFTs in the same growth group are
competing more strongly with each other than those that are in different
growth groups.

Differences to the parameterization of
\citet{Dislich-Simulatingforestdynamics-2009} are particularly evident in the
mortality parameters. Lower values were estimated for the mortality of the
midsuccessional PFTs 3 and 4, while mortality of the late successional PFTs
5, 6 and 7 was estimated higher. This pattern is mirrored in the maximum
diameter growth rates of midsuccessional species. Thus, our study points to
less pronounced differences between mid- and late-successional types than
\citet{Dislich-Simulatingforestdynamics-2009}. We can only speculate about
the reason for these differences. In general, one would think that the
systematic parameter estimation is more reliable than the manual calibration
by \citet{Dislich-Simulatingforestdynamics-2009}. However, although
\citet{Dislich-Simulatingforestdynamics-2009} calibrated to the same data,
they also considered the fit of other model outputs such as total biomass and
expert opinions for fixing the parameters. Expert opinion in particular would
favor more pronounced differences in mortality rates between mid- and
late-successional species due to ecological expectations, although specific
empirical data on tree mortality or on maximum growth rates under full light
were not available. Secondly, there are significant correlations between the
parameters, which allow us to gain a similar fit with a range of different
parameter values. And finally, we were using the model in this study at a
lower temporal resolution (5\,yr time steps) than
\citet{Dislich-Simulatingforestdynamics-2009} to reduce computing time, which
can affect model dynamics and equilibrium distributions, meaning that
slightly different parameter values would be estimated for the same model
with different temporal resolution.

\subsection{Advantages compared to conventional calibration\hack{\newline} methods}

Our results demonstrate that inverse parameter estimation with a likelihood
function derived from the stochasticity in the model outputs is feasible and
provides good results, even for a relatively complex and runtime-intensive
ecological model. This is encouraging in itself, as it is neither trivial to
calibrate a parameter-rich model with heterogeneous data in general, nor easy
to address all the technical challenges for performing the simulation-based
likelihood approximation. A valid question, however, is whether the gain is
worth the effort -- after all, our approach is connected with considerable
computational and conceptual costs, and all we gain are parameter estimates
that could probably also have been derived with conventional inversion
methods such as parameter optimization.

We believe the effort is justified, particularly because there are practical
advantages of simulation-based likelihood approximations for ecological
research that extend far beyond what we could demonstrate in this study.
First of all, there is considerable interest in connecting models to large
and heterogeneous data sources that become increasingly available
\citep{Luo-Ecologicalforecastingand-2011,
Hartig-Connectingdynamicvegetation-2012,
Dietze-improvingcommunicationbetween-2013}. A practical problem in this
context is that conventional methods provide no good answer as to how
different data sources should be weighted to construct a joint likelihood or
objective function. Moreover, ecological processes almost inevitably lead to
correlations between those different data types, meaning that we would not
expect errors to be independent, posing a challenge for conventional methods.
Simulation-based likelihood approximations provide a natural answer to these
problems. Assuming that the simulation model includes all major sources of
stochasticity, likelihoods approximations automatically weight the importance
of different model outputs and account for correlations between them. In our
study, we can see this in the combined fit of growth rates and stem size
distributions, which required no weighting of these two patterns and
automatically accounted for correlations between them.

Moreover, under conventional inverse parameterization procedures, one might
see that a certain pattern is not well represented, but it is often difficult
to decide whether this is a random or a systematic problem. Simulation-based
likelihood approximations allow us to make a definite statement about the
probability of observed patterns given the current model (parameters). Thus,
we can use the full arsenal of statistical procedures, including Bayesian and
frequentist model selection, to compare alternative ecological hypotheses.
The possibility of such rigorous statistical tests for alternative
process-based models will likely increase the acceptance of process-based
models as a tool, not only for representing and predicting but also for
statistically testing ecological knowledge.

\subsection{Differences to ABC}

The comments in the previous subsection apply to parametric and
nonparametric likelihood approximation alike. However, it also seems
interesting to discuss differences between the parametric likelihood
approximation used in this study and the more widely used nonparametric
approximation used in ABC \citep{Beaumont-ApproximateBayesianComputation-2010}. As
discussed in \citet{Hartig-Statisticalinferencestochastic-2011},
unlike ABC, parametric likelihood approximations will almost inevitably
exhibit a certain amount of bias because it is unlikely that a simple
distributional model can emulate model output distributions in all respects
(particularly in the tails of the output distribution). Yet, the parametric
approximation also has practical advantages. Many ecological models have to
be run into equilibrium before predictions can be made. Once such a model is
in equilibrium, more draws for the parametric approximation can be generated
relatively cheaply, while a new run has to be started for each ABC step. In our
example, the time required for the parametric approximation in one MCMC step
was not much longer than for an ABC step, but the parametric approximation
ensures a good acceptance probability. To reach the same acceptance
probability with ABC, we would have to accept a relatively large ABC
approximation error. This error may be corrected later, but the fact remains
that, for situations where the number of possible MCMC evaluation is fixed
(complex models), both ABC and parametric approximations will have a
nonnegligible error. We conjecture that the balance could well be in favor
of parametric approximations in situations such as the one encountered in
this study.

\conclusions

Our results suggest that likelihood approximations, in particular parametric
likelihood approximations, are a promising route for the parameterization of
stochastic ecological models. Their use is technically more challenging than
the ``traditional'' Bayesian approach where likelihoods are based on
phenomenological error models. The advantage, however, is that error models
are based on the same ecological mechanisms as all other model predictions.
Thus, they allow a more rigorous test of the mechanistic model assumptions,
because the mechanisms have to explain both the mean and the variance in the
data. Moreover, likelihood approximations account for the relative importance
and correlations between different data types predicted by the model, which
makes them interesting when models have to be coupled to heterogeneous data.
In this study, additional computational costs of the approach were moderate
(factor 2--5) compared to a standard Bayesian approach due to the fact that
the model had to be run into equilibrium in any case. Such runtime
differences appear secondary compared to the methodological advantage of
rigorously testing our mechanistic understanding of ecosystems against field
data, including the sampling and measurement process. Parametric likelihood
approximations therefore seem particularly promising for models that have to
be run into equilibrium, contain the dominant stochastic processes, use
heterogeneous data, and predict outputs that can be well summarized by
standard distributions.

\hack{\newpage}

\Supplementary{pdf }

\begin{acknowledgements}
We would like to thank Anja~Rammig, Christopher~Reyer and Susanne~Rolinski
for their helpful comments during the review process, and Anne Carney for
proofreading. FH acknowledges support from ERC advanced grant 233066. C.~Dislich was
supported by the German Research Foundation (DFG, Research Unit 816).
 \hack{\newline}
\hack{\newline} The service charges for this open access publication
\hack{\\}have been covered by a Research Centre of the \hack{\\}Helmholtz
Association.
\hack{\newline} \hack{\newline} Edited by: A.~Rammig
\end{acknowledgements}

\end{document}